\documentclass[11pt]{article}

\usepackage{amsmath}
\usepackage{graphicx}
\usepackage{indentfirst}
\usepackage{amssymb}
\usepackage{cite}
\usepackage{color}
\usepackage{subfigure}
\usepackage{varwidth}
\usepackage{subfigure}
\usepackage[colorlinks=true, linkcolor=red, citecolor=blue, urlcolor=magenta]{hyperref}
\usepackage{xcolor}

\begin{document}
\title{Distinguishing Black Holes and Neutron Stars through Optical Images}

\date{}
\maketitle

\begin{center}
\author{Chen-Yu Yang,}$^{a}$\footnote{E-mail: chenyuyang\_2024@163.com}
\author{Xiao-Xiong Zeng}$^{b}$\footnote{E-mail: xxzengphysics@163.com (Corresponding author)}
\\

\vskip 0.25in
$^{a}$\it{Department of Mechanics, Chongqing Jiaotong University, Chongqing 400000, People's Republic of China}\\
$^{b}$\it{College of Physics and Electronic Engineering, Chongqing Normal University, Chongqing 401331, People's Republic of China}\\
\end{center}
\vskip 0.6in
{\abstract
{
This paper employs the backward ray tracing method to study the optical images of neutron stars under the conditions of a spherical light source and a thin accretion disk, considering a polynomial equation of state given by $p = K \rho^{1 + 1/n_c}$. By numerically solving the TOV equations, we obtain the interior solutions of neutron stars for different densities. The results indicate that as the polynomial index $n_c$ increases, the mass, radius, and compactness of the neutron star all increase, which has a significant impact on its optical properties. Under the assumption that the light is truncated at the surface of the neutron star, we find that for a spherical light source, an increase in $n_c$ leads to an enlargement of the Einstein ring radius. For a thin accretion disk, the light intensity always reaches its maximum at the surface of the neutron star. The increase in $n_c$ also causes the outline of the neutron star to grow. When the observer inclination angle $\theta_o$ changes, the neutron star's outline deforms from a circular shape to a D shape, with the left side being significantly brighter than the right side. In addition, this paper also investigates the distribution characteristics of the redshift factor. At lower observer inclination angles, gravitational redshift dominates, while at higher inclination angles, the Doppler effect induces noticeable blueshift. Compared to the Schwarzschild black hole, the optical appearance of the neutron star shows significant differences. The study provides a theoretical basis for distinguishing neutron stars from black holes using high-resolution imaging and for constraining the equation of state.
}}

\thispagestyle{empty}
\newpage
\setcounter{page}{1}

\newpage
\section{Introduction}
Neutron stars are the most compact stars in the universe. A typical neutron star has a mass of about $M \sim 1-2M_{\odot}$ ($M_{\odot} = 2 \times 10^{33} \, \mathrm{g}$ is the solar mass) and a radius of about $R \approx 10-14 \, \mathrm{km}$. The mass density inside a neutron star is approximately $\rho \sim 10^{15} \, \mathrm{g\,cm}^{-3}$, roughly three times the normal nuclear density (i.e., the density of a heavy atomic nucleus), $\rho_0 = 2.8 \times 10^{14} \, \mathrm{g\,cm}^{-3}$. The central density of a neutron star can even be one order of magnitude higher than $\rho_0$ \cite{baym1975neutron, baym1979physics, heiselberg2000recent}. Such extreme conditions cannot be realized in the laboratory, and their specific properties, even the composition, remain highly uncertain. Although neutron stars are named for their primary composition of neutrons, they still contain a small number of protons, as well as enough electrons and mesons to maintain overall electrical neutrality. Under supranuclear densities, strangeness-bearing baryons \cite{glendenning1984neutron}, condensed mesons (such as pions or kaons) \cite{kaplan1986strange, kaplan1988kaon}, and even deconfined quark matter \cite{collins1975superdense} may appear. Currently, multiple theoretical models have been established to describe the internal structure and state of matter within neutron stars, but to confirm the superiority of any particular model, it still requires systematic analysis and interpretation of related observational data. Neutron stars exhibit various unique properties and produce numerous observable phenomena, providing valuable opportunities for testing theoretical models of extreme states of matter \cite{fortov2009extreme}.

Stars lose mass due to material ejection during their evolution. Stars with initial masses of $M < 6-8M_{\odot}$ will eject a large amount of material after going through the red giant phase, eventually evolving into white dwarfs with a mass of about $0.5-0.6M_{\odot}$. White dwarfs resist gravitational collapse through electron degeneracy pressure. However, if the remaining mass exceeds the Chandrasekhar limit (approximately $1.3M_{\odot}$), the electron degeneracy pressure can no longer maintain the equilibrium of the star, and the stellar core undergoes rapid contraction. Ultimately, the star can only be supported by neutron degeneracy pressure, leading to the formation of a neutron star. The formation of a neutron star is usually accompanied by a Type-II, Ib, or Ic supernova event \cite{vidana2018short}.

The study of neutron stars dates back to the 1930s. After Chadwick's discovery of the neutron, Baade and Zwicky first proposed the concept of a neutron star in 1934 \cite{baade1934remarks}, predicting that neutron stars are the products of supernova explosions. In 1939, Oppenheimer and Volkoff provided the first computational results for the neutron star model \cite{oppenheimer1939massive}, namely the famous TOV equations. However, due to the long-term lack of observable physical effects, the concept of neutron stars was largely ignored in the field of astronomy. It wasn't until the discovery of pulsars in late 1967 that Gold suggested pulsars are rapidly rotating neutron stars \cite{gold1979rotating}, a viewpoint that sparked widespread interest. In 1975, Hulse and Taylor first discovered a pulsar in a binary system \cite{hulse1975discovery}, making it possible to determine the physical properties of neutron stars through astronomical observations. By measuring the decay of the orbital period in the binary system, the existence of gravitational waves was indirectly confirmed \cite{taylor1979measurements,weisberg2010timing}. In addition, after accreting companion material, a neutron star may further collapse to form a black hole \cite{potekhin2010physics}.

In 2019, the Event Horizon Telescope (EHT) directly captured the first image of a black hole shadow \cite{akiyama2019first1, akiyama2019first2, akiyama2019first3, akiyama2019first4, akiyama2019first5, akiyama2019first6}. This achievement not only marked the beginning of a new era in black hole shadow research \cite{zeng2025holographic, guo2024image, cui2024optical, guo2024influence, he2024observational, hou2024unique, huang2024images, zhang2024imaging, zeng2020shadows, yang2024shadow, zeng2022shadows}, but also greatly stimulated interest in the optical images of various compact objects \cite{zeng2025optical, he2025optical}. The optical images of black hole shadows and compact objects share many similarities. For example, when a celestial object is illuminated by its own accretion disk, a bright ring-like structure forms, with a region of decreased brightness at the center. These bright rings are formed by photons that are bent by strong gravitational lensing effects \cite{luminet1979image, falcke1999viewing, wambsganss1998gravitational}. This theoretical viewpoint is consistent with the observed image of the supermassive object at the center of the M87 galaxy, surrounded by an over-heated plasma. On the other hand, with respect to optical images (or shadows), different compact objects may exhibit significant differences in the structure of the photon sphere, such as the presence of multiple photon rings or the complete absence of photon rings \cite{wielgus2020reflection, tsukamoto2021gravitational, olmo2022new, guerrero2022light, tsukamoto2022retrolensing}. These structural differences provide an important theoretical basis for distinguishing different compact objects through optical images.

A great deal of research has been conducted on the polynomial form of the equation of state (EOS) for neutron stars. Flanagan and Hinderer were the first to calculate the tidal Love numbers for fully relativistic neutron star models with a polynomial pressure-density relation $P = K \rho^{1 + 1/n_c}$ at $l = 2$ \cite{flanagan2008constraining}. Lattimer and Prakash discussed the reasonable range of values for $n_c$ \cite{lattimer2007neutron}. Hinderer later provided a detailed explanation of the calculation process \cite{hinderer2009erratum}, and Read extended the polynomial model to a piecewise polytrope form \cite{read2009constraints}. This paper will investigate the optical images of neutron stars in different accretion disk backgrounds based on the above equation of state. By adjusting the polynomial index $n_c$, we will explore the effect of density on the optical images.

The structure of this paper is organized as follows. Section \ref{sec2} provides a brief introduction to the construction method of the interior equilibrium equations of neutron stars. Section \ref{sec3} derives the equations of motion for light rays near a neutron star and presents the optical images for a spherical light source. Section \ref{sec4} introduces the imaging method for a thin accretion disk background and provides the numerical results for the optical images and the distribution of the redshift factor. Finally, Section \ref{sec5} offers a brief summary and discussion. This paper adopts geometric units, with $c = G = 1$, where $c$ is the speed of light in a vacuum and $G$ is the gravitational constant.

\section{Equilibrium Configuration}\label{sec2}
The spacetime geometry of a static, spherically symmetric star can be described by the following metric \cite{misner1973gravitation}
\begin{equation}
	ds^2=g_{\mu\nu}dx^\mu dx^\nu=-e^{A(r)}dt^2+e^{B(r)}dr^2+r^2(d\theta^2+\sin^2\theta d\varphi^2).\label{eq:ssm}
\end{equation}
The material field inside the star can be approximated as an ideal fluid, with the stress-energy tensor given by
\begin{equation}
	T_{\mu\nu}=(\rho+p)u_{\mu}u_{\nu}+pg_{\mu\nu},
\end{equation}
where $u^\mu=e^{-A/2} \partial_t$ represents the four-velocity of the fluid, and $\rho$ and $p$ are the energy density and pressure of the fluid, respectively. From Einstein's field equations, the TOV equation can be derived
\begin{equation}
	\frac{dp}{dr}=-(\rho+p)\frac{m(r)+4\pi pr^{3}}{r\left[r-2m(r)\right]},\label{eq:tov}
\end{equation}
where
\begin{align}
	&m(r)=4\pi\int_{0}^{r}\rho(x)x^{2}dx,\\
	&\frac{dm(r)}{dr}=4\pi\rho(r)r^{2},\label{eq:me}\\
	&\frac{dA}{dr}=2\times \frac{m(r)+4\pi pr^{3}}{r[r-2m(r)]}.
\end{align}
In order to numerically integrate equation (\ref{eq:tov}), an equation of state $f(p,\rho)=0$ must be introduced, where $f$ represents a functional form. Generally, the pressure $p$ depends not only on the density $\rho$, but also on the specific entropy (the average entropy per nucleon) and the chemical composition inside the star. Only when the specific entropy and chemical composition are uniform throughout the star, $p$ can be considered a function of $\rho$, and the equation of state can be expressed as $f(p,\rho)=0$. For example, in ordinary stars like the Sun, the specific entropy is not constant, whereas in compact objects such as white dwarfs and neutron stars, the specific entropy is typically assumed to be zero everywhere \cite{liang2023differential}. In this paper, we will adopt a polynomial form for the equation of state \cite{hinderer2009erratum, read2009constraints}
\begin{equation}
	p=K\rho^\varGamma,
\end{equation}
where $K$ is a constant and $\varGamma$ is the adiabatic index, given by
\begin{equation}
	\varGamma=1+\frac{1}{n_c},
\end{equation}
where $n_c$ is the polynomial index. By solving the TOV equation (\ref{eq:tov}), the radius $R$ and total mass (energy) $M=m(R)$ of the neutron star can be obtained, where $M$ includes the contribution from gravitational potential energy. The numerical solution for the metric inside the star can also be determined, and in the external region of the star, the Schwarzschild metric can be used. The two metrics must satisfy the junction condition at the surface of the star
\begin{equation}
	e^{A(R)}=1-\frac{2M}{R}.
\end{equation}

It should be noted that the interior of a neutron star is a high-density nuclear matter region. Therefore, we assume that photons are completely absorbed when they reach the surface of the neutron star. This paper mainly focuses on the effect of spacetime curvature on the propagation of light and the characteristics of imaging, and does not address the specific details of radiation transfer and absorption mechanisms. Thus, in the following, the stellar radius $R$ is treated as the optical opacity boundary, meaning that the propagation of light is terminated when it reaches this radius, without considering reflection or re-emission phenomena. This truncation condition reflects both the complete opacity of the neutron star's surface to electromagnetic radiation and, in numerical calculations, avoids non-physical integrals arising from light rays in high-curvature regions, thus ensuring the stability of the imaging computations.

For the convenience of applying the backward ray-tracing techniques \cite{cunha2015shadows}, this paper adopts smooth analytical functions to uniformly fit the metric components inside and outside the star. The fitting goal is to minimize the error as much as possible while ensuring that the fitted metric asymptotically approaches the Schwarzschild metric as $r \to \infty$. After trials, it was found that the following two functions fit the metric components well
\begin{align}
	F_t & = -\exp\left[a_7\left(\exp\left(-\frac{1+a_1r+a_2r^2}{a_3+a_4r+a_5r^2+a_6r^3}\right)-1\right)\right],\label{eq:fit1}\\
	F_r &= \exp\left[b_7 \left( \exp\left( -\frac{1 + b_1 r + b_2 r^2}{b_3 + b_4 r + b_5 r^2 + b_6 r^3} \right) - 1 \right) \right].\label{eq:fit2}
\end{align}
Based on these fitting functions, the optical images of neutron stars under different light source conditions can be simulated.

\section{Optical Images under a Spherical Light Source}\label{sec3}
This section will explore the optical images of neutron stars under a spherical light source. The foundation for achieving this goal is to derive the equations of motion for light rays near a neutron star. For convenience in subsequent discussions, we rewrite the metric (\ref{eq:ssm}) as
\begin{equation}
	ds^2=-\mathcal{A}(r)dt^2+\mathcal{B}(r)^{-1}dr^2+r^2d\theta^2+r^2\sin^2\theta d\varphi^2.
\end{equation}
The photon’s trajectory satisfies the Euler-Lagrange equation
\begin{equation}
	\frac{d}{d\lambda}\left(\frac{\partial L}{\partial\dot{x}^\mu}\right)=\frac{\partial L}{\partial x^\mu},
\end{equation}
where $\dot{x}^{\mu}$ is the photon’s four-velocity, the symbol "$\cdot$" denotes differentiation with respect to an affine parameter $\lambda$ along the null geodesic, and $L$ is the Lagrangian. Let $x^\mu=\{t,r,\theta,\varphi\}$, then the Lagrangian can be expressed as
\begin{align}
	0=L&=-\frac{1}{2}g_{\mu\nu}\dot{x}^{\mu}\dot{x}^{\nu} \notag\\
	&=-\frac{1}{2}\left[-\mathcal{A}(r)\dot{t}^{2}+\frac{1}{\mathcal{B}(r)}\dot{r}^{2}+r^{2}\dot{\theta}^{2}+r^{2}\sin^{2}\theta\dot{\varphi}^{2}\right].\label{eq:lag}
\end{align}
Since the metric components do not explicitly depend on $t$ and $\varphi$, the spacetime possesses two Killing vector fields, $\xi_{t}=\partial_{t}$ and $\xi_{\varphi}=\partial_{\varphi}$, corresponding to time translation symmetry and rotational symmetry, respectively. Based on this, we restrict the null geodesic to the equatorial plane, i.e., $\theta=\pi/2$, which introduces two conserved quantities
\begin{align}
	\mathcal{E} &= \frac{\partial L}{\partial\dot{t}}=\mathcal{A}(r)\frac{dt}{d\lambda},\label{eq:e}\\
	\mathcal{L} &= -\frac{\partial L}{\partial\dot{\varphi}}=r^2\frac{d\varphi}{d\lambda}.\label{eq:l}
\end{align}
By combining equations (\ref{eq:lag})-(\ref{eq:l}), we obtain the components of the photon’s four-velocity
\begin{align}
	\dot{t} &= \frac{1}{\mathcal{I}\mathcal{A}(r)}, \label{eq:geo1} \\
	\dot{r} &= \sqrt{\frac{1}{\mathcal{I}^2}\frac{\mathcal{B}(r)}{\mathcal{A}(r)} - \frac{1}{r^2}\mathcal{B}(r)}, \\
	\dot{\theta} &= 0, \\
	\dot{\varphi} &= \pm\frac{1}{r^2}. \label{eq:geo4}
\end{align}
The symbol "$+$" represents clockwise propagation, and "$-$" represents counterclockwise propagation. The impact parameter $\mathcal{I}$ is defined as
\begin{equation}
	\mathcal{I} \equiv \frac{{|\mathcal{L}|}}{\mathcal{E}}.\label{impara}
\end{equation}

Equations (\ref{eq:geo1})-(\ref{eq:geo4}) constitute the first-order differential equations for the photon geodesics. To determine the photon’s trajectory, the corresponding integration constants must also be specified, which requires the choice of an observer. Given the arbitrariness in the observer's spacetime position, we select a zero-angular-momentum observer (ZAMO). Suppose the observer is located at $(t_o, r_o, \theta_o, \varphi_o)$, and within its neighborhood, there exists a locally orthonormal tetrad
\begin{align}
	\tilde{e}_0 = \frac{1}{\sqrt{-g_{tt}}}\partial_t,\quad \tilde{e}_1 = -\frac{1}{\sqrt{g_{rr}}}\partial_r, \nonumber\\
	\tilde{e}_2 = \frac{1}{\sqrt{g_{\theta\theta}}}\partial_{\theta},\quad \tilde{e}_3 = -\frac{1}{\sqrt{g_{\varphi\varphi}}}\partial_{\varphi}.
	\label{eq:frame}
\end{align}
It should be noted that the choice of frame is not unique, and different frames are related by Lorentz transformations. 

To describe the direction of the light ray from the observer’s perspective, we introduce celestial coordinates. Let $\overrightarrow{OM}$ be the projection of the tangent vector of the photon geodesic at point $O$ onto the observer's three-dimensional subspace, i.e., the photon’s three-momentum. Define $\Theta_{1}$ as the angle between $\overrightarrow{OM}$ and $\tilde{e}_1$, and $\Theta_{2}$ as the angle between $\overrightarrow{OA}$ and $\tilde{e}_2$. The celestial coordinates are represented by the coordinate system $(\Theta_{1}, \Theta_{2})$. In the local frame (\ref{eq:frame}), the tangent vector of the photon geodesic can be expressed as
\begin{align}
	\dot{S} = |\overrightarrow{OM}|\left(-\tilde{e}_{0} + \tilde{e}_{1}\cos\Theta_{1} + \tilde{e}_{2}\sin\Theta_{1}\cos\Theta_{2} + \tilde{e}_{3}\sin\Theta_{1}\sin\Theta_{2}\right).
\end{align}
The negative sign in front of $\tilde{e}_{0}$ ensures that the tangent vector points in the direction of the past null geodesic. Additionally, since the photon’s trajectory is independent of its energy, we set the photon energy to $1$, i.e., $|\overrightarrow{OM}|=1$.

On the other hand, for each light ray, the coordinates $(t, r, \theta, \varphi)$ are functions of the affine parameter $\lambda$. The general form of the tangent vector can be expressed as
\begin{equation}
	\dot{S} = \dot{t}\,\partial_t + \dot{r}\,\partial_r + \dot{\theta}\,\partial_\theta + \dot{\varphi}\,\partial_\varphi. \label{eq:ltv}
\end{equation}
From equations (\ref{eq:frame})-(\ref{eq:ltv}), it can be seen that there is a one-to-one correspondence between the photon’s four-momentum and the celestial coordinates. Once the photon’s four-momentum is given, the corresponding celestial coordinates can be uniquely determined. Conversely, if the celestial coordinates are known, the corresponding four-momentum can be obtained through coordinate transformation. Therefore, by combining the observer’s location, the initial conditions of the photon’s equations of motion can be determined.

To obtain the optical image of the neutron star, this paper adopts the fisheye camera model, which maps the celestial coordinates $(\Theta_{1}, \Theta_{2})$ to Cartesian coordinates $(X, Y)$ on the imaging plane. The main advantage of this model is its wide field of view. The projection coordinates of point $M$ on the imaging plane are given by
\begin{equation}
	X_M = -2|\overrightarrow{OM}|\tan\left(\frac{\Theta_{1}}{2}\right)\sin\Theta_{2},\quad Y_M = -2|\overrightarrow{OM}|\tan\left(\frac{\Theta_{1}}{2}\right)\cos\Theta_{2}. \label{eq:proco1}
\end{equation}
The size of the imaging plane is determined by the field of view angle $\Phi_{\mathrm{fov}}$. We take $\Phi_{\mathrm{fov}}/2$ in both the $X$ and $Y$ directions, defining a square screen with a side length of
\begin{equation}
	l = 2\left|\overrightarrow{OM}\right|\tan\frac{\Phi_{\mathrm{fov}}}{2}.
\end{equation}
The imaging plane is divided into $m \times m$ pixels, and the side length of each pixel is
\begin{equation}
	\ell = \frac{l}{2} = \frac{2}{m}\left|\overrightarrow{OM}\right|\tan\frac{\Phi_{\mathrm{fov}}}{2}.
\end{equation}
Each pixel is labeled by coordinates $(a, b)$, where the bottom-left pixel is $(1, 1)$ and the top-right pixel is $(m, m)$, with $1 \leqslant a,b \leqslant m$. The relationship between $(X_M, Y_M)$ and the pixel coordinates $(a, b)$ is given by
\begin{equation}
	X_M = \ell\left(a - \frac{m+1}{2}\right),\quad Y_M = \ell\left(b - \frac{m+1}{2}\right). \label{eq:proco2}
\end{equation}
By comparing equations (\ref{eq:proco1}) and (\ref{eq:proco2}), the correspondence between the celestial coordinates $(\Theta_{1}, \Theta_{2})$ and the pixel coordinates $(a, b)$ is obtained as
\begin{align}
	\Theta_{1} &= 2\,\arctan\left[\frac{1}{m}\tan\left(\frac{\Phi_{\mathrm{fov}}}{2}\right)\sqrt{\left(a - \frac{m+1}{2}\right)^2 + \left(b - \frac{m+1}{2}\right)^2}\right], \\ 
	\Theta_{2} &= \arctan\left[\frac{2b - (m+1)}{2a - (m+1)}\right].
\end{align}
The above analysis lays the foundation for the implementation of the backward ray-tracing method. Using the reversibility of light propagation, the backward ray-tracing method assumes that photons are emitted by the observer and the imaging is completed by numerically integrating the photon geodesic equations. The advantage of this method is that it does not require considering the rays that start from the light source but fail to reach the observer, thus significantly simplifying the computational process. Once a photon reaches the celestial sphere, the corresponding pixel's color is uniquely determined.

\begin{figure}[!h]
	\centering 
	\subfigure{\includegraphics[scale=0.5]{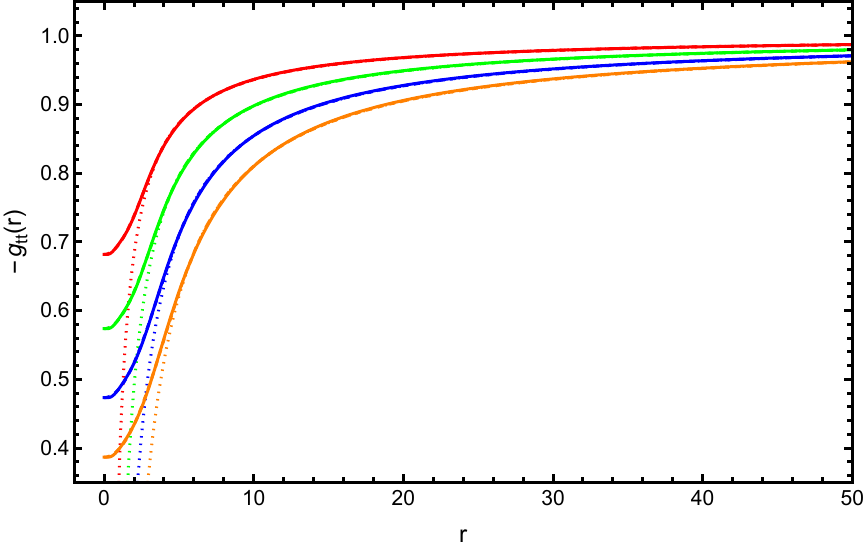}}
	\subfigure{\includegraphics[scale=0.5]{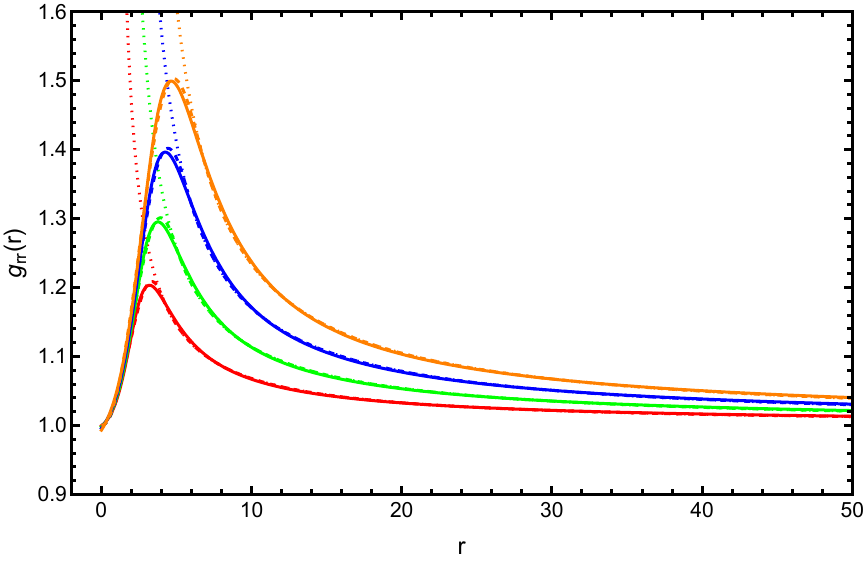}}
	
	\caption{Metric components corresponding to different polynomial indices $n_c$. The left figure shows $-g_{tt}$, and the right figure shows $g_{rr}$. The red, green, blue, and orange curves correspond to $n_c = 1.1, 1.2, 1.3, 1.4$, respectively. The solid and dashed lines represent the fitted metric and numerical metric, while the dotted line represents the Schwarzschild metric for the mass $M$ (neutron star mass).}\label{fig1}
\end{figure}

Figure \ref{fig1} shows the numerical results of the metric components $-g_{tt}$ (left) and $g_{rr}$ (right). The red, green, blue, and orange curves correspond to $n_c = 1.1, 1.2, 1.3, 1.4$, respectively. The solid line represents the fitted metric, the dashed line represents the numerical metric, and the dotted line represents the Schwarzschild metric corresponding to the mass $M$ (neutron star mass). As can be seen from the figure, for both $-g_{tt}$ and $g_{rr}$, the numerical results almost completely coincide with the fitted results (for relative error analysis, see Appendix \ref{appendix1}). Therefore, the fitted metric can be used to replace the numerical metric for subsequent calculations. Since the neutron star does not have an event horizon, the metric components do not diverge as $r \to 0$. As $r \to \infty$, the fitted metric and the Schwarzschild metric exhibit the same asymptotic behavior, indicating that this is an asymptotically flat spacetime.

\begin{figure}[!h]
	\centering 
	\subfigure[$n_c=1.1$]{\includegraphics[scale=0.35]{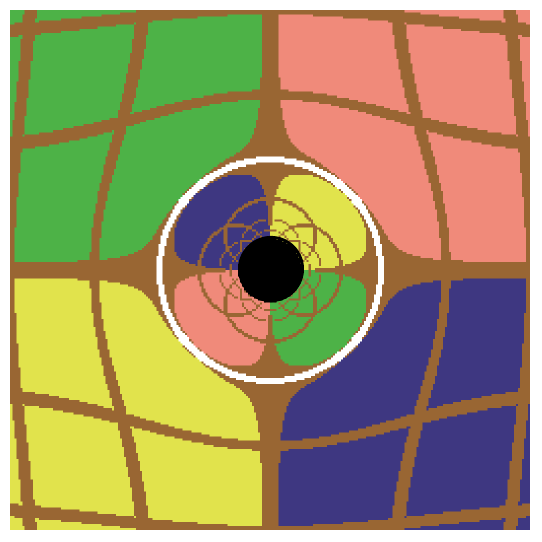}}
	\subfigure[$n_c=1.2$]{\includegraphics[scale=0.35]{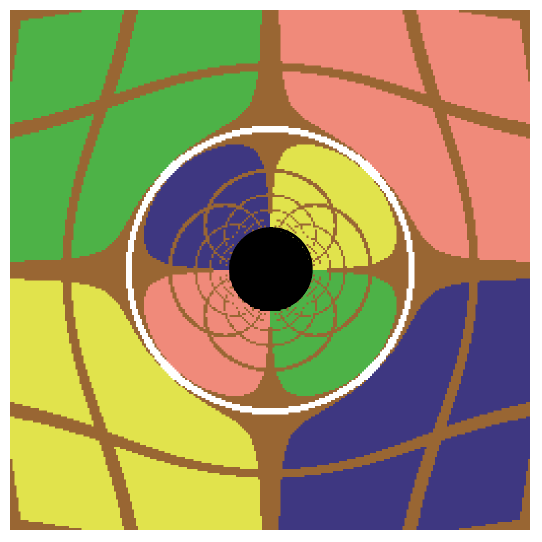}}
	\subfigure[$n_c=1.3$]{\includegraphics[scale=0.35]{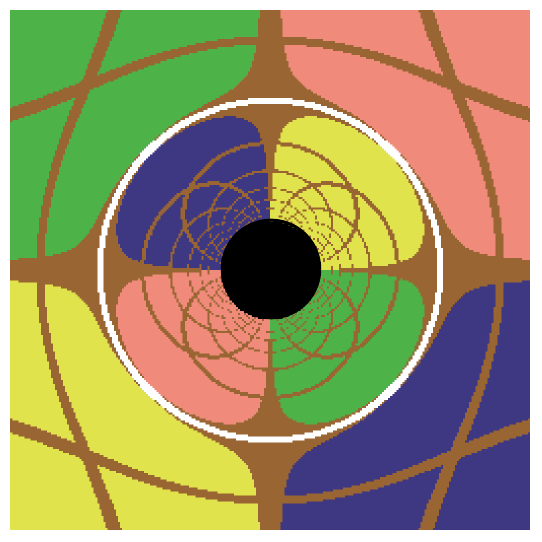}}
	\subfigure[$n_c=1.4$]{\includegraphics[scale=0.35]{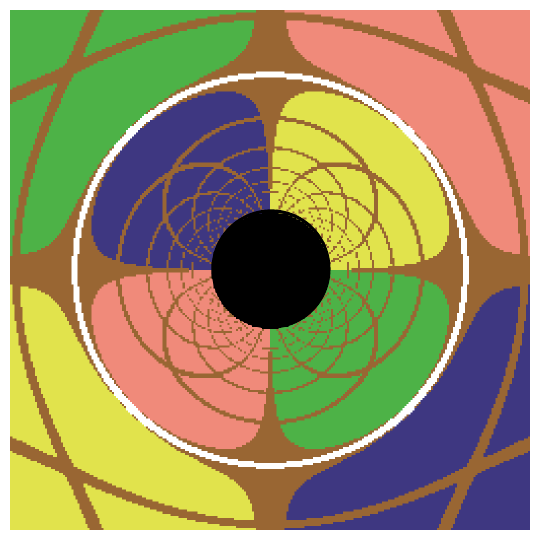}}
	
	\caption{Optical image of the neutron star under a spherical light source. The black region represents the neutron star, while the white ring represents the Einstein ring. The fixed parameters are observer inclination angle $\theta_o = 45^\circ$, observer distance $r_o = 200$, and field of view angle $\Phi_{\mathrm{fov}} = 20^\circ$.}\label{fig2}
\end{figure}

Tables \ref{tab1} and \ref{tab2} present the relevant parameters of the neutron star and the estimated values of the fitting functions. Figure \ref{fig2} shows the optical image of the neutron star under a spherical light source. The parameter settings are the observer inclination angle $\theta_o = 45^\circ$, the observer distance $r_o = 200$, and the field of view angle $\Phi_{\mathrm{fov}} = 20^\circ$. To enhance the visual effect of the image, the celestial sphere is divided into four quadrants, which are marked in red, blue, yellow, and green, respectively. The neutron star is located at the center of the celestial sphere, while the observer is at the intersection of the four quadrants on the celestial sphere's surface. On the celestial sphere, a white reference light source is placed at the observer's antipode to study the Einstein ring phenomenon caused by strong gravitational lensing effects. The black region at the center of the image represents the outline of the neutron star. As mentioned earlier, the light is truncated at the neutron star's radius, forming a shadow structure similar to that of a black hole. The white outer ring represents the Einstein ring, with the brown dashed line inside representing the photon trajectory.

Several interesting phenomena can be observed from the figure. For all the images, due to the use of the spherically symmetric metric (\ref{eq:ssm}), which does not include a rotation parameter, the optical image of the neutron star in the imaging plane always appears as a standard circle. The background celestial sphere does not exhibit any dragging effects, similar to the shadow of a static black hole. Additionally, the Einstein ring in the image remains a complete closed circle without any fragmentation.

The optical image of the neutron star is primarily determined by the equation of state. To investigate this, we studied the effect of the polynomial index $n_c$ on the imaging results. Figure \ref{fig2} shows the cases for $n_c = 1.1, 1.2, 1.3, 1.4$ from left to right. As $n_c$ increases, both the radius $R$ and mass $M$ of the neutron star increase. Therefore, the more physically meaningful parameter is the compactness $C = M/R$. As shown in Table \ref{tab1}, $C$ increases with $n_c$, indicating that the neutron star becomes denser, and the degree of spacetime curvature also increases. This is reflected in the image by the significantly larger neutron star radius and Einstein ring radius.

\begin{table}[ht]
	\centering
	\caption{Estimated values of the parameters $a_i$ in the fitting function $F_t$ for $g_{tt}$, where $M$ and $R$ represent the mass and radius of the neutron star, and $C = M/R$ is the compactness.}
	\footnotesize
	\begin{tabular}{c|c|c|c|c|c|c|c|c|c|c}
		\hline
		$n_c$ & $M$ & $R$ & $C$ & $a_1$ & $a_2$ & $a_3$ & $a_4$ & $a_5$ & $a_6$ & $a_7$ \\ \hline
		1.1 & 0.319 & 4.012 & 0.079 & -0.316 & 0.201 & 0.018 & 0.475 & -0.299 & 0.121 & 0.383 \\ \hline
		1.2 & 0.511 & 4.916 & 0.104 & -0.237 & 0.135 & 0.022 & 0.406 & -0.210 & 0.073 & 0.555 \\ \hline
		1.3 & 0.727 & 5.823 & 0.125 & -0.181 & 0.098 & 0.028 & 0.359 & -0.159 & 0.050 & 0.748 \\ \hline
		1.4 & 0.948 & 6.756 & 0.140 & -0.139 & 0.075 & 0.036 & 0.324 & -0.125 & 0.037 & 0.950 \\ \hline
	\end{tabular}
	\label{tab1}
\end{table}

\begin{table}[ht]
	\centering
	\caption{Estimated values of the parameters $b_i$ in the fitting function $F_r$ for $g_{rr}$.}
	\footnotesize
	\begin{tabular}{c|c|c|c|c|c|c|c}
		\hline
		$n_c$ & $b_1$ & $b_2$ & $b_3$ & $b_4$ & $b_5$ & $b_6$ & $b_7$ \\ \hline
		1.1 & -37.684 & -22.136 & $7.737\times10^{7}$ & $-3.220\times10^{7}$ & $5.002\times10^{6}$ & 962915.956 & 30528.399 \\ \hline
		1.2 & -11.392 & -4.329 & $3.818\times10^{7}$ & $-1.352\times10^{7}$ & $1.863\times10^{6}$ & 266240.230 & 69894.293 \\ \hline
		1.3 & -6.898 & -1.798 & $1.462\times10^{7}$ & $-4.578\times10^{6}$ & 583700.455 & 64746.066 & 58661.396 \\ \hline
		1.4 & -5.106 & -0.956 & $1.419\times10^{7}$ & $-4.031\times10^{6}$ & 492175.440 & 44043.006 & 98090.056 \\ \hline
	\end{tabular}
	\label{tab2}
\end{table}

\section{Optical Images and Redshift Factor Distribution under a Thin Accretion Disk}\label{sec4}
An isolated neutron star will eventually exhaust its thermal and magnetic energy, gradually fading and disappearing. However, if the neutron star is in a binary system, the situation becomes more complex. For instance, the neutron star can accrete a large amount of material from its companion star, and the released gravitational and thermal nuclear energy will make it a bright X-ray source \cite{potekhin2010physics, lipunov1992astrophysics}. In this case, the accreted material will form an accretion disk around the neutron star. The accretion disk itself also emits X-rays, and due to the precession of the accretion disk or changes in the accretion rate, the luminosity of the neutron star will vary with time. If the amount of accreted material exceeds a certain critical value, the neutron star will collapse and eventually form a black hole. Overall, due to the presence of the accretion disk, the optical image of the neutron star will change significantly. Therefore, studying the optical images of neutron stars illuminated by an accretion disk as a light source has important physical significance. For simplicity, we assume that the optical and geometrical thin accretion disk lies in the equatorial plane, and the observer is sufficiently far away, i.e., $r_o \gg M$. Further, we neglect the refraction effects of light during propagation. Under these assumptions, when the light interacts with the accretion disk, the radiation intensity changes due to photon emission and absorption. The non-polarized radiation transfer equation is then
\begin{equation}
	\frac{d}{d\lambda}\left(\frac{\mathcal{S}_\nu}{\nu^3}\right) = \frac{e_\nu - a_\nu \mathcal{S}_\nu}{\nu^2},
\end{equation}
where $\lambda$ is the photon’s affine parameter, and $\mathcal{S}_{\nu}$, $e_{\nu}$, and $a_{\nu}$ represent the specific intensity, emissivity, and absorption coefficient at frequency $\nu$, respectively. In the thin accretion disk approximation, only the instantaneous emission and absorption of photons in the equatorial plane need to be considered, so outside the equatorial plane, $e_{\nu}=a_{\nu}=0$. At this point, $\mathcal{S}_{\nu}/\nu^3$ is conserved along the null geodesics, and the total intensity on the observer’s screen is given by
\begin{equation}
	\mathcal{S}_o = \sum_{n=1}^{N} f_n e_n (g_n)^3,
\end{equation}
where $n=1,\ldots,N$ represents the number of times the photon passes through the equatorial plane, and $f_n$, $e_n$, and $g_n = \nu_o / \nu_n$ are the fudge factor, emissivity, and redshift factor, respectively. When $n=1$, a direct image is formed; when $n=2$, a lensed image is formed; and when $n \geqslant 2$, higher-order images are formed.

From the above equation, it can be seen that the intensity $\mathcal{S}_o$ is determined by $f_n$, $e_n$, and $g_n$. Below is a brief discussion on the values of these three factors. The fudge factor $f_n$ describes the absorption characteristics of the accretion disk and needs to be determined based on a specific accretion disk model. Existing studies have shown that $f_n$ primarily affects the intensity of the photon ring and has a limited impact on the overall optical image. Therefore, in this paper, we set $f_n = 1$ \cite{he2025optical}. For different accretion disk models, the emissivity $e_n$ has different choices. To remain consistent with astronomical observations (such as the imaging results of M87$^{\star}$ and Sgr A$^{\star}$), $e_n$ is typically assumed to be a second-order polynomial in log-space. For example, in the Gralla-Lupsasca-Marrone model \cite{gralla2020shape},
\begin{equation}
	e_n = \frac{\exp\left[-\frac{1}{2}\left(c_1\operatorname{arcsinh}\left(\frac{r-c_2}{c_3}\right)\right)^2\right]}{\sqrt{(r-c_2)^2+(c_3)^2}},
\end{equation}
where $c_1$, $c_2$, and $c_3$ correspond to the rate of increase, radial translation, and dilation of the profile, respectively. This model is consistent with the accretion disk structure based on general relativistic magnetohydrodynamics (GRMHD) simulations, and it is widely used \cite{vincent2022images}. These parameters together control the spatial distribution of radiated photons, and in principle, they can be adjusted to match numerical simulation results with actual observations. In this paper, we choose $c_1 = 0$, $c_2 = 6M$, and $c_3 = M$, where $M$ is the mass of the neutron star. 

The redshift factor $g_n$ has a significant impact on the intensity $\mathcal{S}_o$ \cite{luis2022shadows, rosa2023imaging}. $g_n$ includes contributions from both Doppler redshift and gravitational redshift. Among these, Doppler redshift plays a dominant role in explaining the distribution of intensity in the image, while gravitational redshift reflects the effect of strong gravitational fields on the photon propagation path. Assuming the accreted material is electrically neutral plasma, moving along null geodesics, and similar to equations (\ref{eq:e}) and (\ref{eq:l}), this geodesic also has two conserved quantities. If the accreted material is too close to the neutron star, it will fall into the surface of the star; if it is sufficiently far, it may move along a stable circular orbit. For black holes, the boundary orbit between these two scenarios is called the innermost stable circular orbit (ISCO), and we adopt this terminology in this paper. Outside the ISCO, the angular velocity of the fluid can be expressed as
\begin{equation}
	\tilde{\omega}_n = \left.\frac{u^\varphi}{u^t}\right|_{r=r_n},
\end{equation}
where $r_n$ is the radial coordinate when the photon passes through the equatorial plane for the $n$-th time. The redshift factor $g_n$ can be rewritten as
\begin{equation}
	g_n = -\frac{k}{K(1-\mathcal{I}\tilde{\omega}_n)},
\end{equation}
where $\mathcal{I}$ is the photon impact parameter defined in equation (\ref{impara}), and $k$ is the ratio of the energy observed on the screen to $\mathcal{E}$ defined in equation (\ref{eq:e}). For asymptotically flat spacetime, when the observer is at infinity, $k=1$. $K$ is defined as
\begin{equation}
	K \equiv \left.\sqrt{\frac{1}{g_{tt}+g_{\varphi\varphi}\tilde{\omega }_{n}^{2}}}\right|_{r=r_n}.
\end{equation}
Through the above analysis, we can quantitatively calculate the intensity and redshift factor corresponding to each pixel.

\begin{figure}[!h]
	\centering 
	\subfigure[$\theta_o=0^\circ,n_c=1.1$]{\includegraphics[scale=0.35]{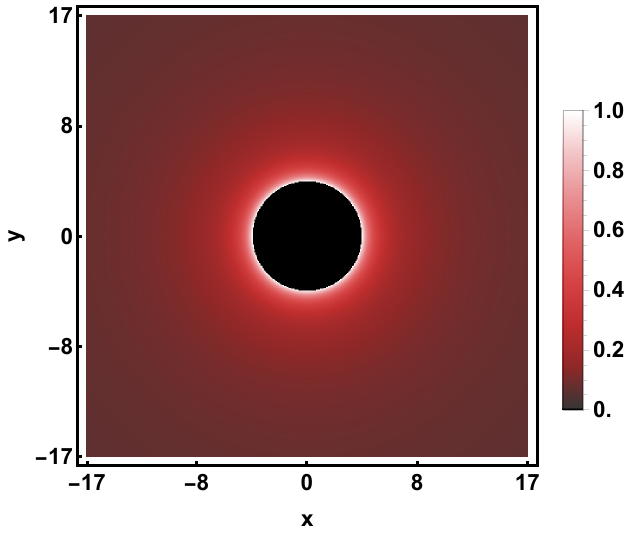}}
	\subfigure[$\theta_o=17^\circ,n_c=1.1$]{\includegraphics[scale=0.35]{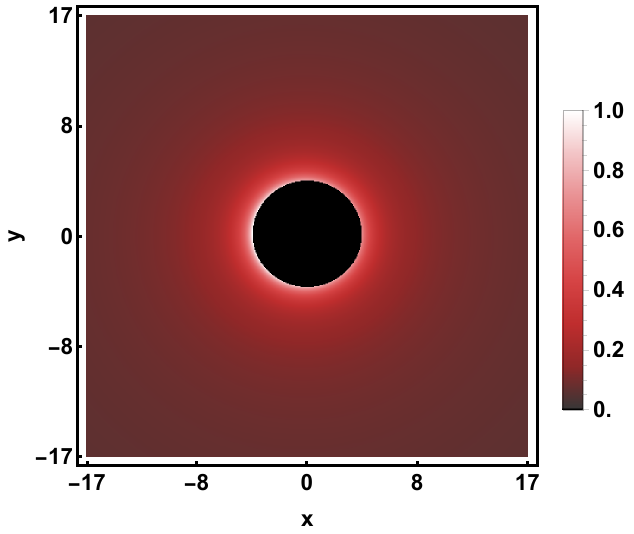}}
	\subfigure[$\theta_o=60^\circ,n_c=1.1$]{\includegraphics[scale=0.35]{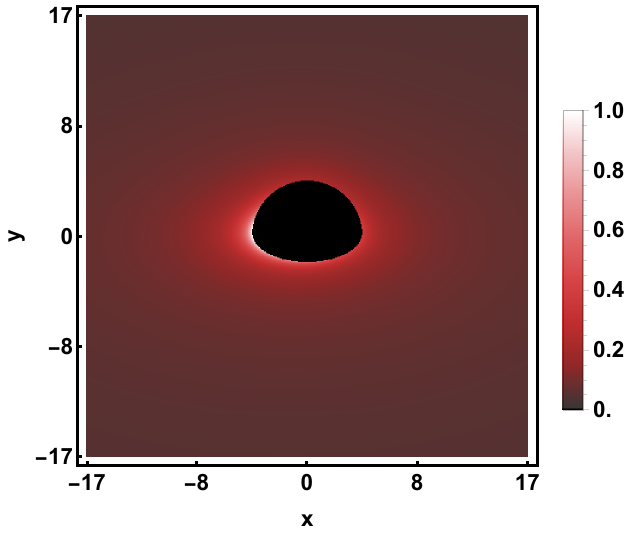}}
	\subfigure[$\theta_o=80^\circ,n_c=1.1$]{\includegraphics[scale=0.35]{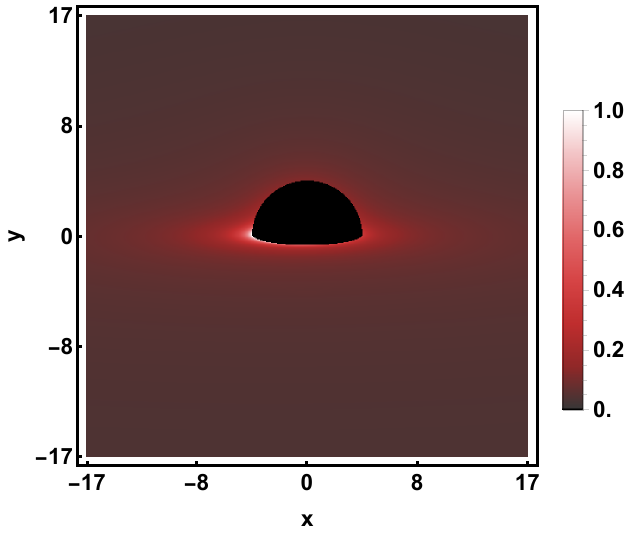}}
	
	\subfigure[$\theta_o=0^\circ,n_c=1.2$]{\includegraphics[scale=0.35]{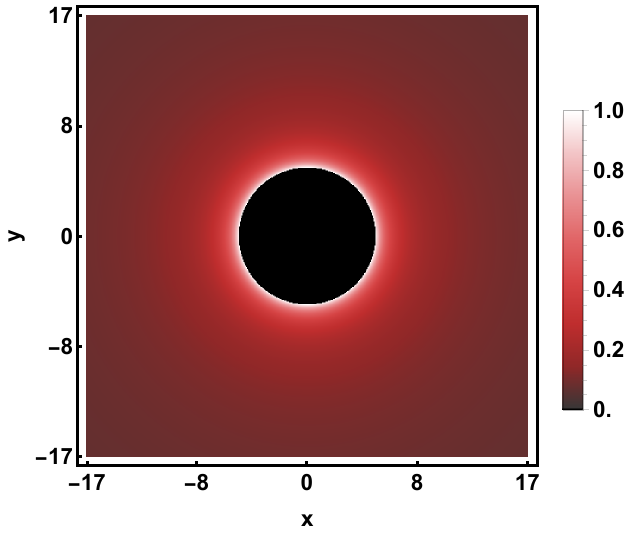}}
	\subfigure[$\theta_o=17^\circ,n_c=1.2$]{\includegraphics[scale=0.35]{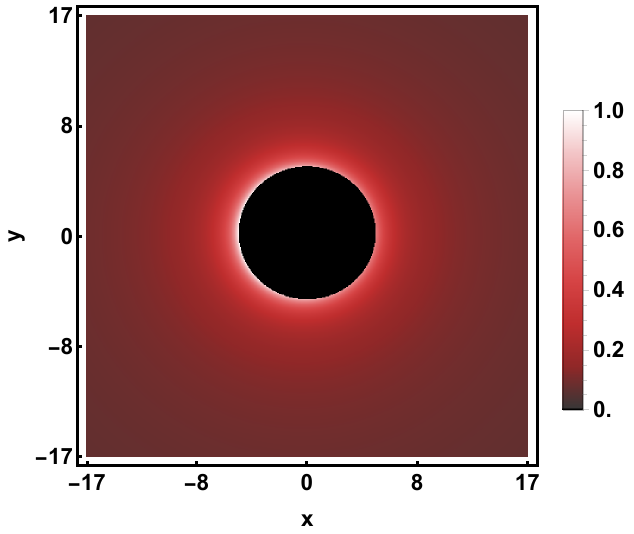}}
	\subfigure[$\theta_o=60^\circ,n_c=1.2$]{\includegraphics[scale=0.35]{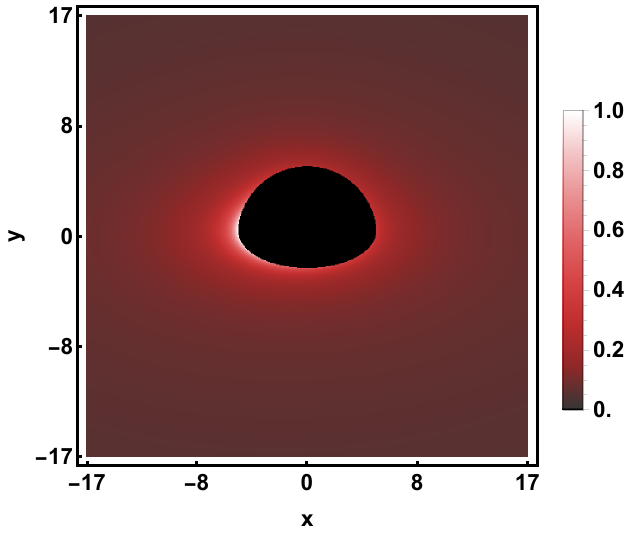}}
	\subfigure[$\theta_o=80^\circ,n_c=1.2$]{\includegraphics[scale=0.35]{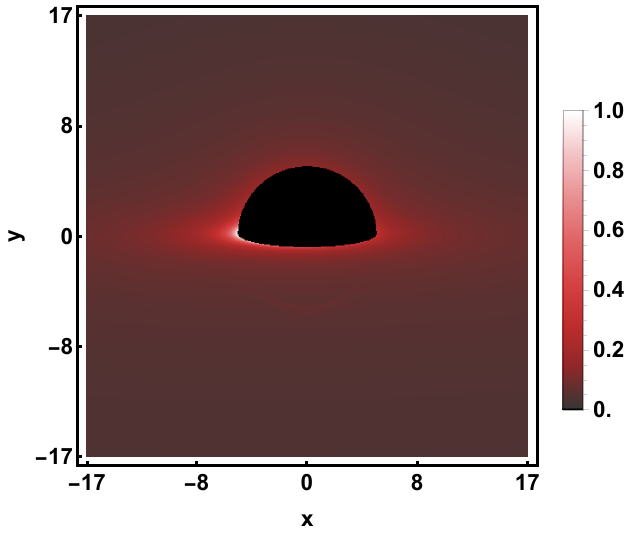}}
	
	\subfigure[$\theta_o=0^\circ,n_c=1.3$]{\includegraphics[scale=0.35]{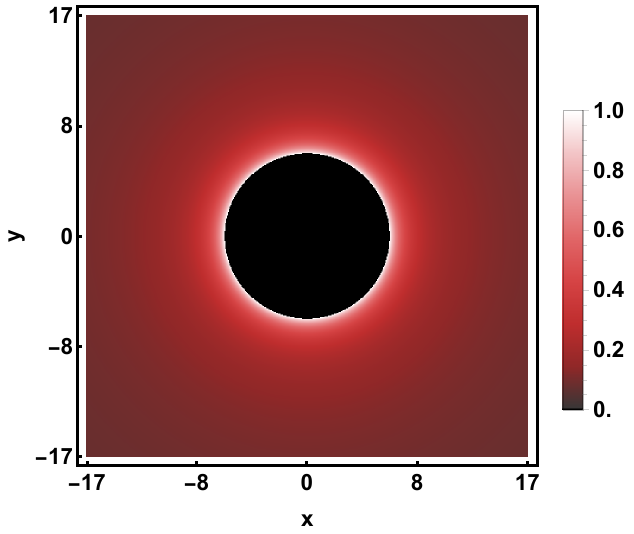}}
	\subfigure[$\theta_o=17^\circ,n_c=1.3$]{\includegraphics[scale=0.35]{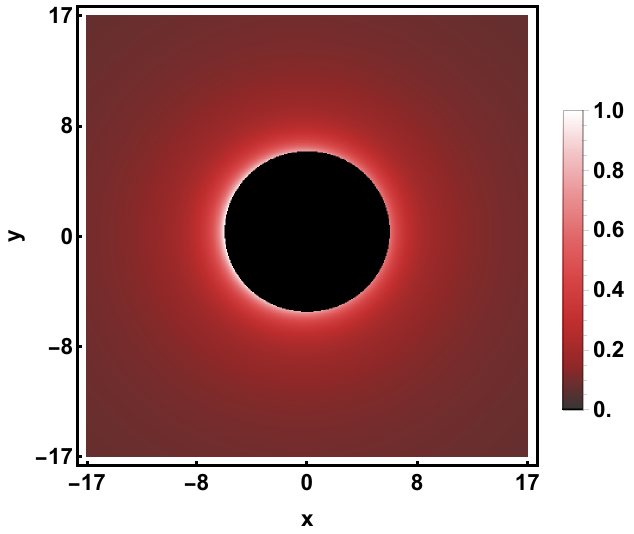}}
	\subfigure[$\theta_o=60^\circ,n_c=1.3$]{\includegraphics[scale=0.35]{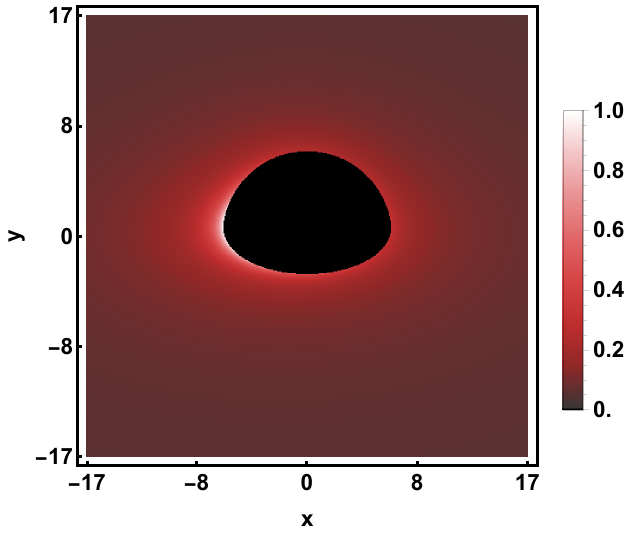}}
	\subfigure[$\theta_o=80^\circ,n_c=1.3$]{\includegraphics[scale=0.35]{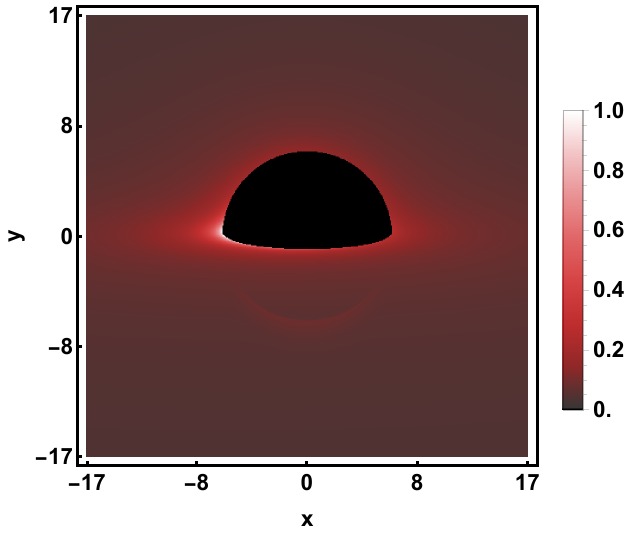}}
	
	\subfigure[$\theta_o=0^\circ,n_c=1.4$]{\includegraphics[scale=0.35]{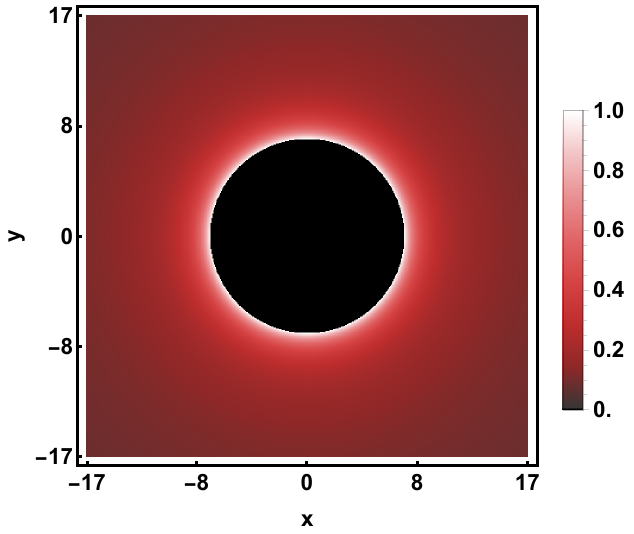}}
	\subfigure[$\theta_o=17^\circ,n_c=1.4$]{\includegraphics[scale=0.35]{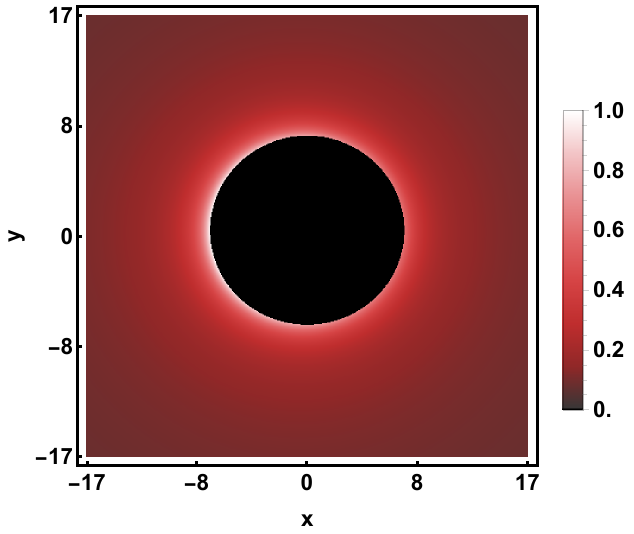}}
	\subfigure[$\theta_o=60^\circ,n_c=1.4$]{\includegraphics[scale=0.35]{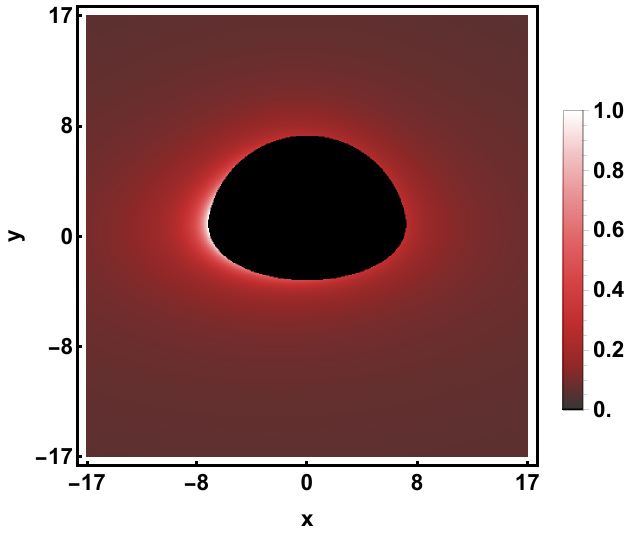}}
	\subfigure[$\theta_o=80^\circ,n_c=1.4$]{\includegraphics[scale=0.35]{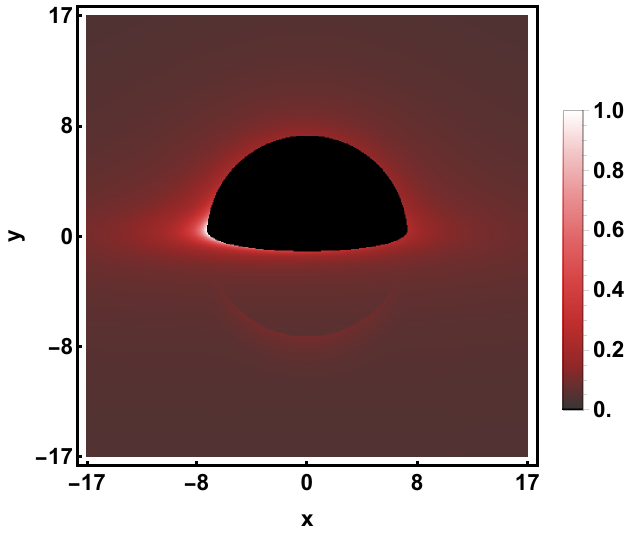}}
	
	\caption{Optical images of the neutron star under a thin accretion disk. From top to bottom, the polynomial indices are $n_c = 1.1, 1.2, 1.3, 1.4$, and from left to right, the observer inclination angles are $\theta_o = 0^\circ, 17^\circ, 60^\circ, 80^\circ$. The fixed parameters are observer distance $r_o = 200$ and field of view angle $\Phi_{\mathrm{fov}} = 10^\circ$.}\label{fig3}
\end{figure}

Figure \ref{fig3} shows the effect of the observer inclination angle $\theta_o$ and the polynomial index $n_c$ on the optical image of the neutron star, with the field of view angle fixed at $\Phi_{\mathrm{fov}} = 10^\circ$ and the observer distance at $r_o = 200$. The neutron star is located at the center of the image, and the colorbar displays the variation in intensity, with white indicating the maximum intensity. Since it is assumed that light is truncated at the surface of the neutron star, the interior of the neutron star is shown in black. From the observations of all the images, it can be seen that the intensity reaches its maximum near the surface of the neutron star, and decreases rapidly as the distance from the star increases. When $\theta_o = 0^\circ$, the neutron star appears as a perfect circle. As $\theta_o$ increases to $17^\circ$, the neutron star undergoes slight deformation. At this point, the intensity on the left side of the star is slightly greater than on the right side due to the Doppler redshift enhancement caused by relative motion. As $\theta_o$ further increases to $80^\circ$, the neutron star's outline deforms into a "D" shape, with the intensity on the left side significantly greater than on the right. Comparing each row, it can be observed that an increase in $n_c$ leads to a larger outline of the neutron star.

\begin{figure}[!h]
	\centering 
	\subfigure[$\theta_o=0^\circ$]{\includegraphics[scale=0.35]{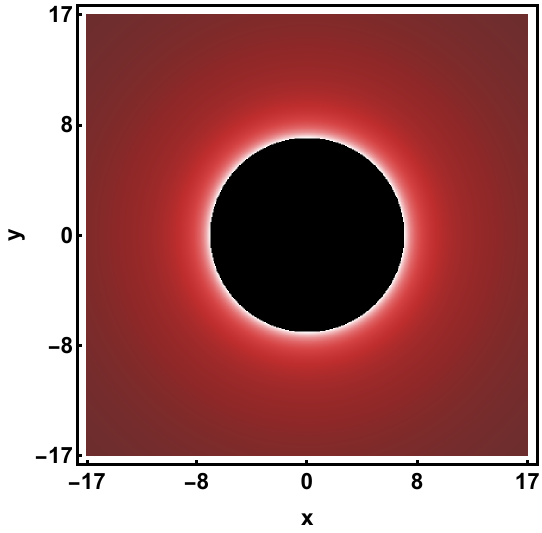}\label{fig44a}}
	\subfigure[$\theta_o=0^\circ$]{\includegraphics[scale=0.35]{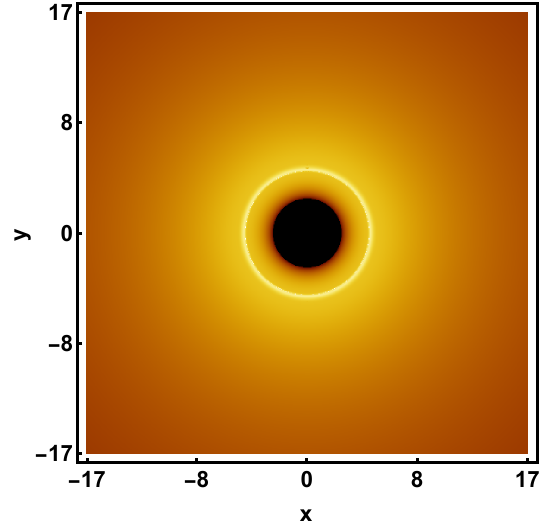}\label{fig44b}}
	\subfigure[$\theta_o=80^\circ$]{\includegraphics[scale=0.35]{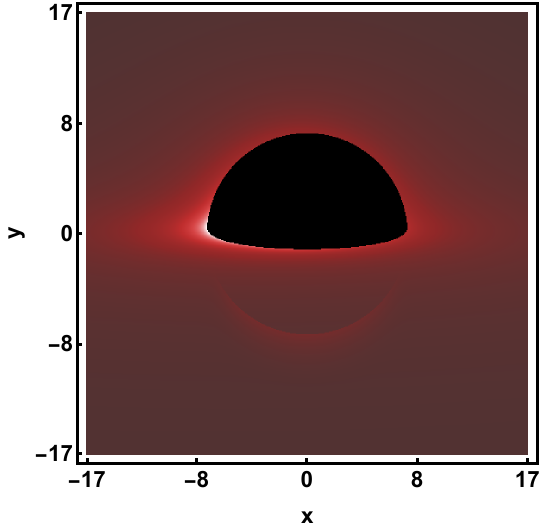}\label{fig44c}}
	\subfigure[$\theta_o=80^\circ$]{\includegraphics[scale=0.35]{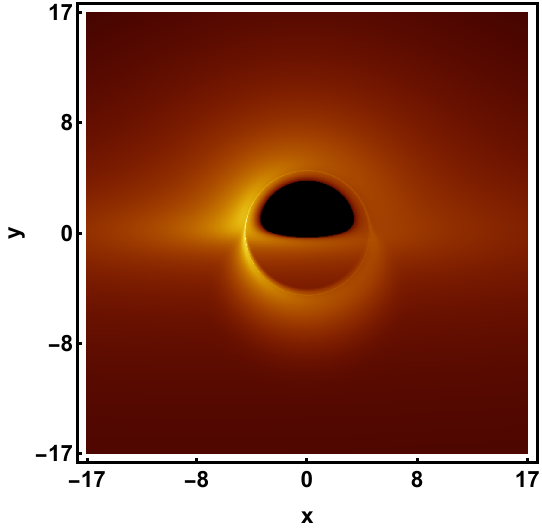}\label{fig44d}}
	
	\caption{Comparison of optical images between the neutron star and the Schwarzschild black hole. Figures \ref{fig44a} and \ref{fig44c} correspond to the neutron star with $n_c = 1.4$. Figures \ref{fig44b} and \ref{fig44d} correspond to the Schwarzschild black hole. For all images, the fixed parameters are mass $M = 0.948$, observer distance $r_o = 200$, and field of view angle $\Phi_{\mathrm{fov}} = 10^\circ$.}\label{fig44}
\end{figure}

To compare the optical images of the neutron star and the Schwarzschild black hole, we plot their images under a thin accretion disk in Figure \ref{fig44}. The black hole has an event horizon at its center, from which photons cannot escape, resulting in a well-defined inner shadow. Under our assumptions, the neutron star also forms a central dark region. Although the exterior of the neutron star uses the Schwarzschild metric, there are significant differences between the two images. For the Schwarzschild black hole, at low observer inclination angles, the region of maximum intensity appears at the photon ring; for the neutron star, the region of maximum intensity appears at the surface of the star. At high observer inclination angles, a clear lensed image appears beneath the Schwarzschild black hole. Although both objects have the same mass, the black hole is more compact, resulting in a smaller central dark region.

\begin{figure}[!h]
	\centering 
	\subfigure[$\theta_o=0^\circ,n_c=1.1$]{\includegraphics[scale=0.35]{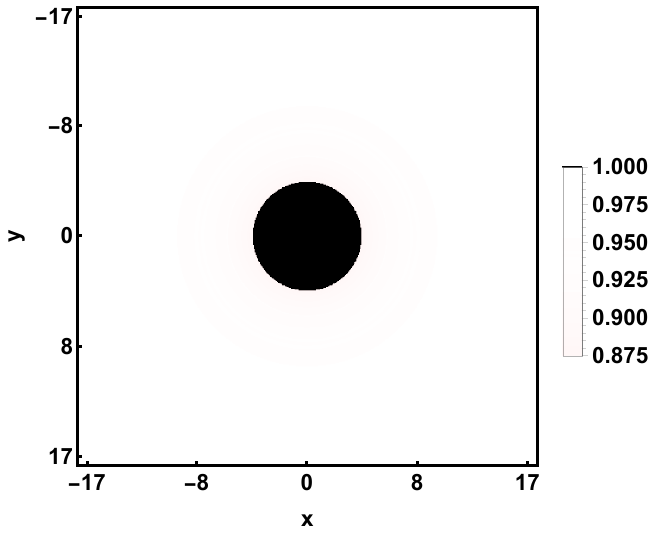}}
	\subfigure[$\theta_o=17^\circ,n_c=1.1$]{\includegraphics[scale=0.35]{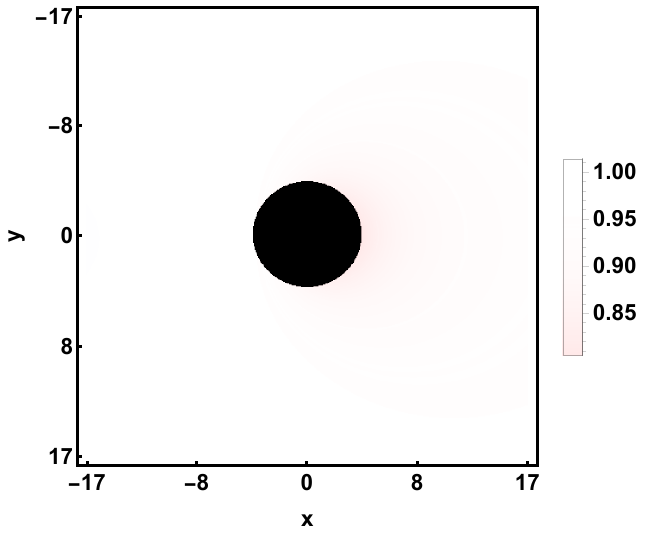}}
	\subfigure[$\theta_o=60^\circ,n_c=1.1$]{\includegraphics[scale=0.35]{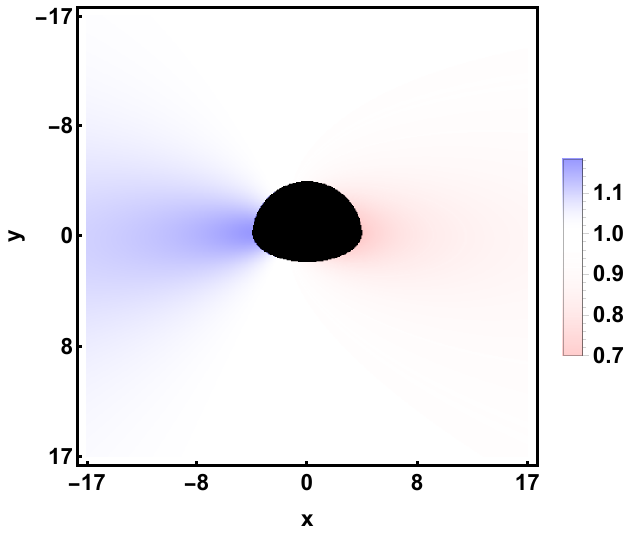}}
	\subfigure[$\theta_o=80^\circ,n_c=1.1$]{\includegraphics[scale=0.35]{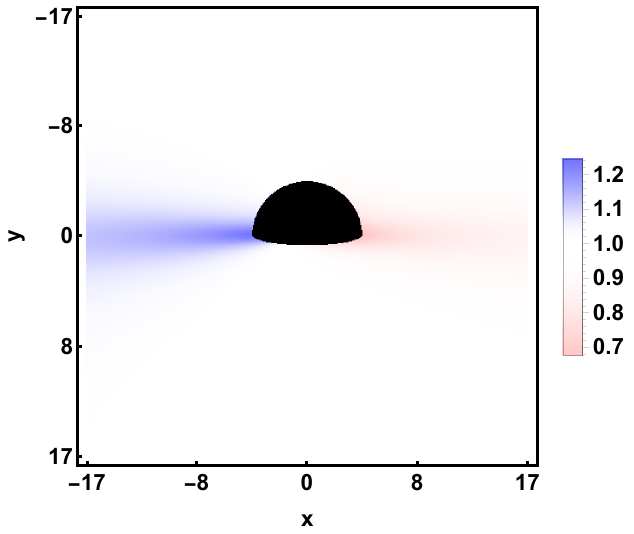}}
	
	\subfigure[$\theta_o=0^\circ,n_c=1.2$]{\includegraphics[scale=0.35]{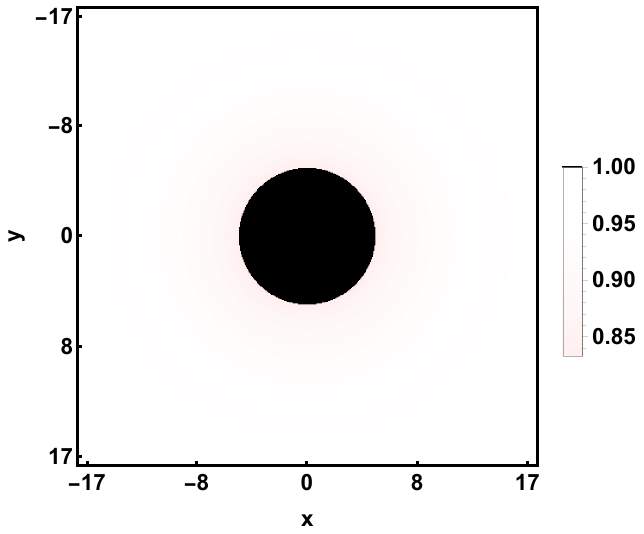}}
	\subfigure[$\theta_o=17^\circ,n_c=1.2$]{\includegraphics[scale=0.35]{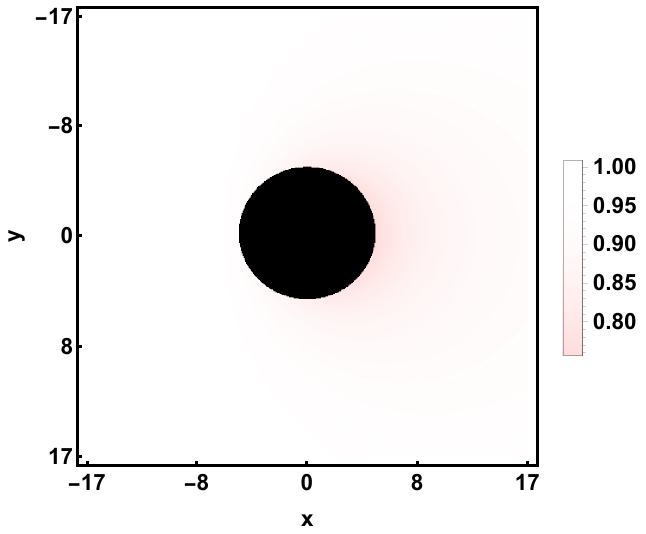}}
	\subfigure[$\theta_o=60^\circ,n_c=1.2$]{\includegraphics[scale=0.35]{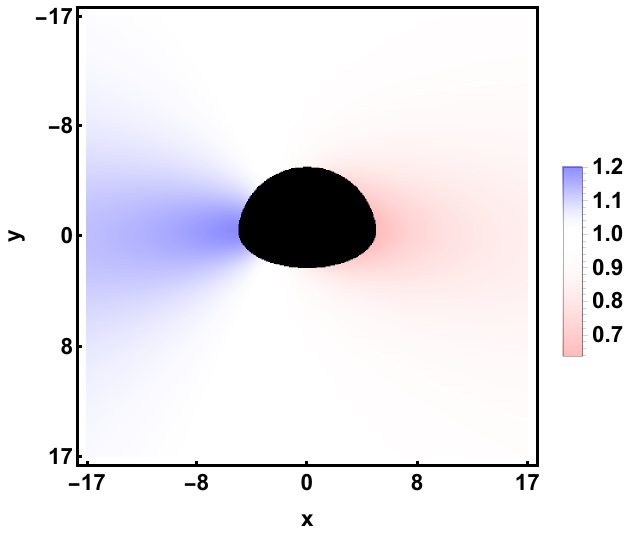}}
	\subfigure[$\theta_o=80^\circ,n_c=1.2$]{\includegraphics[scale=0.35]{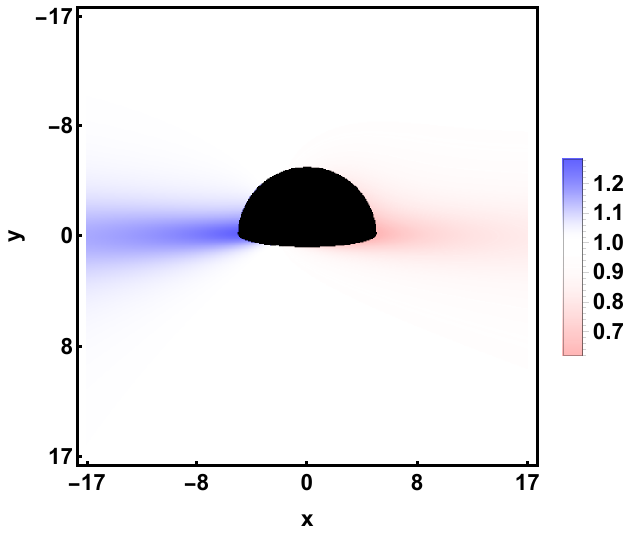}}
	
	\subfigure[$\theta_o=0^\circ,n_c=1.3$]{\includegraphics[scale=0.35]{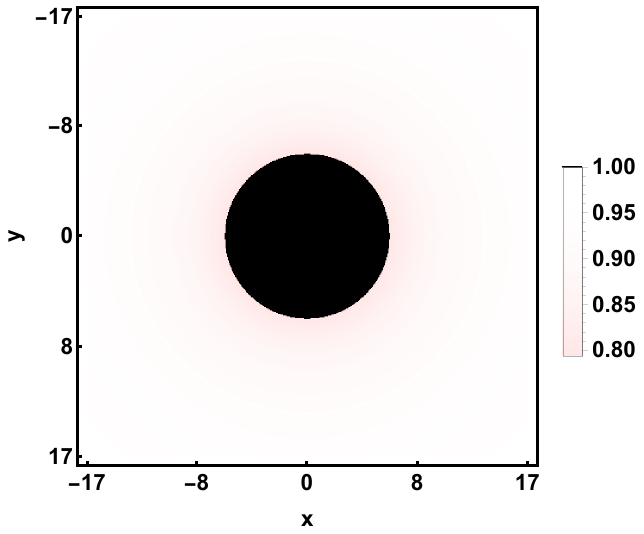}}
	\subfigure[$\theta_o=17^\circ,n_c=1.3$]{\includegraphics[scale=0.35]{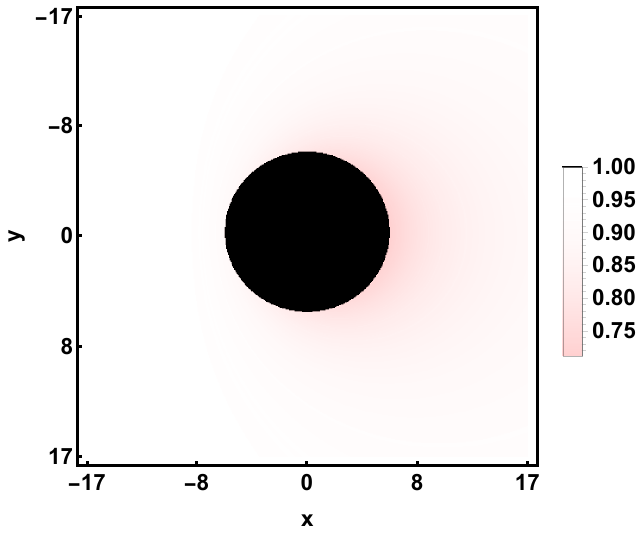}}
	\subfigure[$\theta_o=60^\circ,n_c=1.3$]{\includegraphics[scale=0.35]{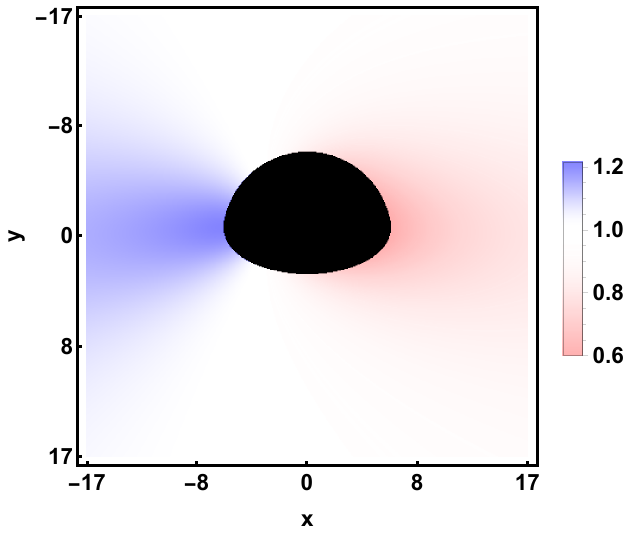}}
	\subfigure[$\theta_o=80^\circ,n_c=1.3$]{\includegraphics[scale=0.35]{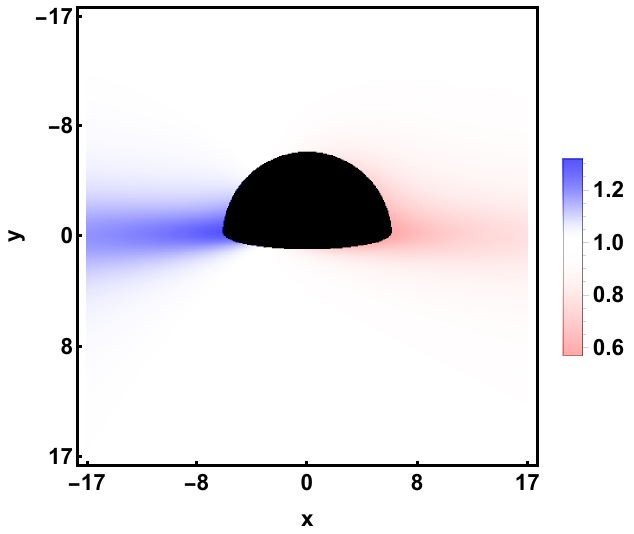}}
	
	\subfigure[$\theta_o=0^\circ,n_c=1.4$]{\includegraphics[scale=0.35]{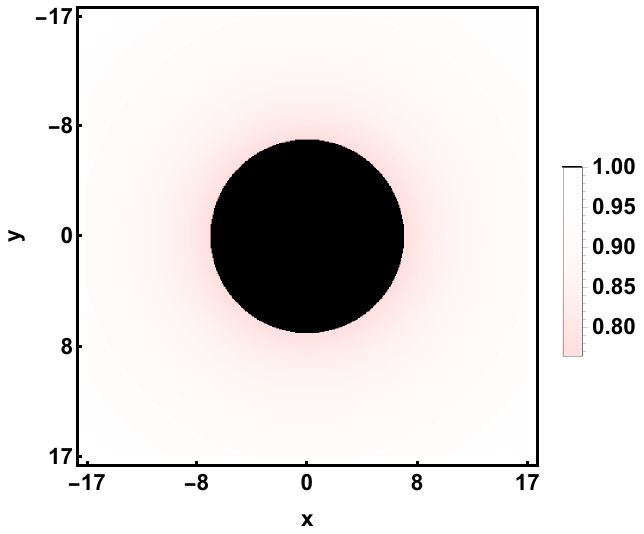}}
	\subfigure[$\theta_o=17^\circ,n_c=1.4$]{\includegraphics[scale=0.35]{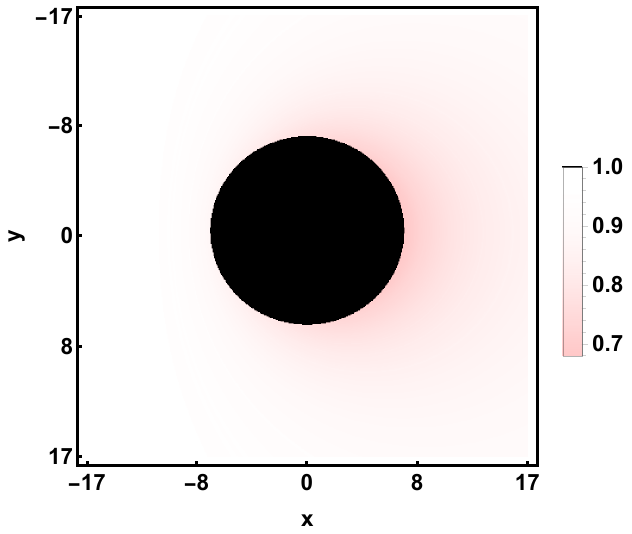}}
	\subfigure[$\theta_o=60^\circ,n_c=1.4$]{\includegraphics[scale=0.35]{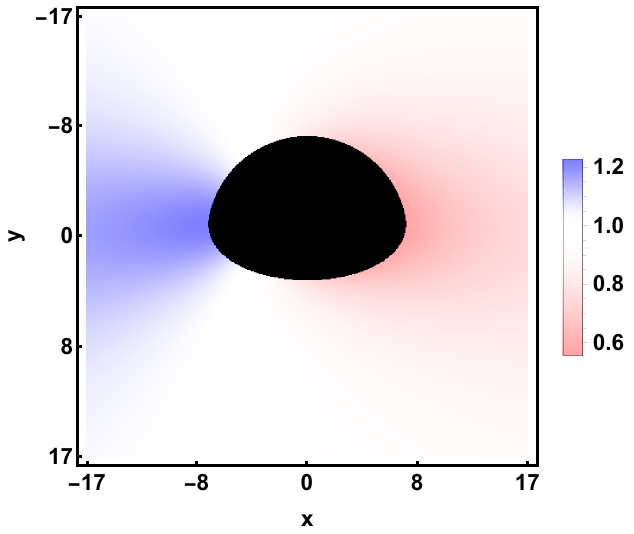}}
	\subfigure[$\theta_o=80^\circ,n_c=1.4$]{\includegraphics[scale=0.35]{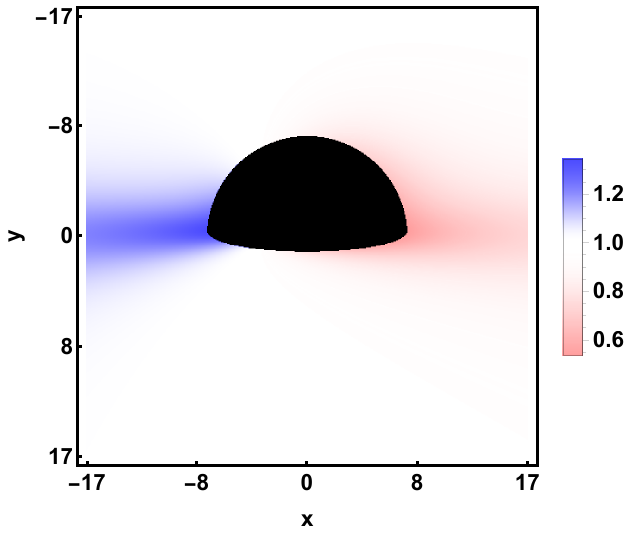}}
	
	\caption{Distribution of the redshift factor for the neutron star under a thin accretion disk. In the figure, red represents redshift and blue represents blueshift. The intensity of the color reflects the strength of the effect, with a linear relationship between the two. From top to bottom, the polynomial indices are $n_c = 1.1, 1.2, 1.3, 1.4$, and from left to right, the observer inclination angles are $\theta_o = 0^\circ, 17^\circ, 60^\circ, 80^\circ$. The fixed parameters are observer distance $r_o = 200$ and field of view angle $\Phi_{\mathrm{fov}} = 10^\circ$.}\label{fig4}
\end{figure}

Figure \ref{fig4} shows the distribution of the redshift factor $g_1$, with the parameter settings consistent with those in Figure \ref{fig3}. In the figure, red represents redshift, and blue represents blueshift. The intensity of the color reflects the strength of the effect, with a linear relationship between the two. From the image, it can be observed that at smaller observer inclination angles (the first and second columns), only redshift is present, with no blueshift. At this point, the redshift primarily comes from gravitational redshift. As $\theta_o$ increases (the third and fourth columns), Doppler redshift becomes more pronounced. A clear blueshift region appears on the left side of these images, corresponding to the phenomenon in Figure \ref{fig3} where the intensity on the left side is greater than on the right side. The increase in $n_c$ concentrates both the redshift and blueshift factors, which is achieved through changes in the compactness $C$.

\section{Concluding Remarks}\label{sec5}
In astronomical observations, the optical images of compact objects may exhibit similar features to the black hole shadow. Therefore, this paper investigates the optical properties of neutron stars under different light source conditions and explores the differences between these and black hole shadows. We consider spherical light sources and thin accretion disks, and adopt a polynomial form for the equation of state. By adjusting the polynomial index $n_c$ and numerically solving the TOV equations, we obtain the interior solutions of neutron stars for different compactnesses. Subsequently, using the backward ray-tracing method, we systematically analyze the effects of $n_c$ and the observer inclination angle $\theta_o$ on the optical images. During the imaging process, we assume that the light is truncated at the surface of the neutron star. Although the neutron star does not have an event horizon, photons can theoretically pass through the surface of the star, but this paper focuses on studying the gravitational effects and neglects phenomena such as refraction, reflection, and absorption of light at the star's surface. Although this truncation means that the intensity is only contributed by the surface or external light sources, slightly lower than the brightness in the case without truncation, the resulting image remains distinctly different from a black hole shadow. This difference is sufficient to reflect the variations between neutron star and black hole imaging.

The results show that as the polynomial index $n_c$ increases, both the mass $M$ and radius $R$ of the neutron star increase. Therefore, in this paper, we introduce the compactness $C = M/R$ for analysis. The calculation results indicate that the compactness $C$ is positively correlated with $n_c$, reflecting the physical characteristic that the neutron star becomes denser as $n_c$ increases. This change has a particularly significant impact on the optical image. Under the spherical light source condition, as $n_c$ increases, both the outline of the neutron star in the imaging plane and the radius of the Einstein ring increase noticeably.

Next, we discussed the optical images under geometric and optical thin accretion disks. For both black holes and neutron stars, regions with decreased intensity appear at the center. However, for black holes, this is due to the event horizon, whereas for neutron stars, it is due to the truncation of the light. Moreover, the images of the two objects show distinct differences. For black holes, the intensity is maximized at the photon ring, while for neutron stars, it is maximized at the surface of the star. Regarding the redshift factor, we found that the observer inclination angle $\theta_o$ has a significant impact on its distribution. When $\theta_o = 0^\circ$ and $17^\circ$, only redshift is present, with gravitational redshift dominating. When $\theta_o = 60^\circ$ and $80^\circ$, blueshift begins to appear, with Doppler redshift significantly enhanced. The increase in the polynomial index $n_c$ makes the distribution of both redshift and blueshift factors more concentrated.

The study of optical images and redshift factors provides a new possibility for constraining the equation of state of neutron stars and helps distinguish neutron stars from black holes in high-resolution astronomical observations. In future research, we will further consider different equations of state and more complex, realistic accretion disk models. These efforts are expected to deepen our understanding of the properties of compact objects.


\cleardoublepage

\vspace{10pt}
\noindent {\bf Acknowledgments}

\noindent
This work is supported by the National Natural Science Foundation of China (Grants Nos. 12375043, 12575069).

\bibliographystyle{JHEP} 
\bibliography{biblio} 

\appendix 

\section{Relative Error Analysis of the Fitted Metric}\label{appendix1}

\begin{figure}[!h]
	\centering 
	\subfigure{\includegraphics[scale=0.5]{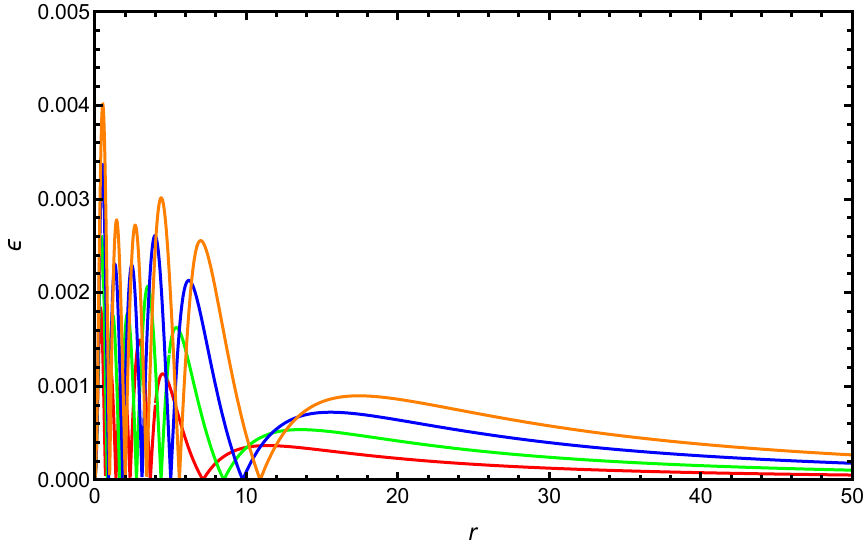}}
	\subfigure{\includegraphics[scale=0.5]{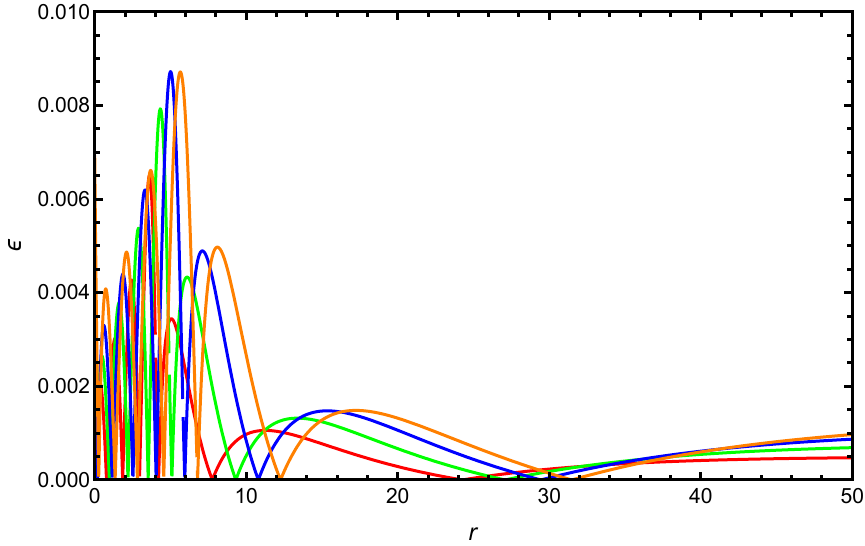}}
	
	\caption{Relative error plots for $-g_{tt}$ (left) and $g_{rr}$ (right). The red, green, blue, and orange correspond to polynomial indices $n_c = 1.1, 1.2, 1.3, 1.4$, respectively.}\label{fig5}
\end{figure}

In Section \ref{sec3}, we calculated the numerical and fitted metrics for the neutron star corresponding to different polynomial indices $n_c$ and used the fitted metric to replace the numerical metric for subsequent calculations. The relative error between the two can be calculated by the following equation
\begin{equation}
	\epsilon=\left|\frac{g_\mathrm{num}-F_\mathrm{fit}}{g_\mathrm{num}}\right|,
\end{equation}
where $g_\mathrm{num}$ is the numerical metric ($-g_{tt}$ or $g_{rr}$), and $F_\mathrm{fit}$ is the fitted metric calculated using equations (\ref{eq:fit1}) and (\ref{eq:fit2}). In Figure \ref{fig5}, we present the relative error graphs for $-g_{tt}$ and $g_{rr}$. The results show that the relative error for $-g_{tt}$ (left figure) is always less than $0.4\%$, while the relative error for $g_{rr}$ (right figure) is always less than $1\%$, and as $r \to \infty$, $\epsilon \to 0$. This indicates that using the fitted metric to replace the numerical metric for calculations is reasonable.

\end{document}